\documentclass[aip,jcp,amsmath,amssymb,reprint]{revtex4-1}
\usepackage{graphicx}
\usepackage{bm}

\begin{document}
\title{
Multiple length and time scales of dynamic heterogeneities in
model glass-forming liquids:
A systematic analysis of multi-point and multi-time correlations
}

\author{Kang Kim}
\email{kin@ims.ac.jp}
\author{Shinji Saito}
\email{shinji@ims.ac.jp}
\affiliation{
Institute for Molecular Science, Okazaki,  Aichi 444-8585, Japan
}
\affiliation{
School of Physical Sciences,
The Graduate University for Advanced Studies, Okazaki, Aichi 444-8585,
Japan
}

\date{\today}

\begin{abstract}
We report an extensive and systematic investigation of the multi-point
 and multi-time correlation functions
to reveal the spatio-temporal structures of dynamic
heterogeneities in glass-forming liquids.
Molecular dynamics simulations are carried out for the supercooled
 states of various prototype
 models of glass-forming liquids such as binary Kob--Andersen,
 Wahnstr\"{o}m, soft-sphere, and network-forming liquids.
While the first three models act as fragile liquids exhibiting
 super-Arrhenius temperature dependence in their relaxation times, the last
 is a strong glass-former exhibiting Arrhenius behavior.
First, we quantify the length scale of the
 dynamic heterogeneities utilizing the four-point correlation function.
The growth of the dynamic length scale with decreasing temperature
 is characterized by various scaling relations that are
 analogous to the critical phenomena.
We also examine how the growth of the length scale depends upon
 the model employed.
Second, the four-point correlation function is extended to a three-time correlation
 function to characterize the temporal structures of the
 dynamic heterogeneities based on our previous studies [K.~Kim
 and S.~Saito, Phys. Rev. E \textbf{79}, 060501(R) (2009); J. Chem. Phys. \textbf{133},
 044511 (2010)].
We provide comprehensive numerical results obtained from the
three-time correlation function for the above models.
From these calculations, we examine the time scale of the dynamic
heterogeneities and determine the associated lifetime in a consistent and
 systematic way.
Our results indicate that the lifetime of the dynamical
heterogeneities becomes much longer than the $\alpha$-relaxation time
 determined from a two-point correlation function in fragile liquids.
The decoupling between the two time scales is remarkable, particularly in
 supercooled states, and the time scales differ by more than an order of
 magnitude in a more fragile liquid.
In contrast, the lifetime is shorter than the $\alpha$-relaxation
 time in tetrahedral network-forming strong liquid, even at lower temperatures.
\end{abstract}

\maketitle

\section{introduction}

Various liquids form disordered and amorphous solids if
temperatures are reduced below their melting points while avoiding crystallizations.
This transition to a disordered solid is known as the glass
transition~\cite{Debenedetti1996Metastable, Donth2001The,
Binder2005Glassy, Wolynes2012Structural}.
A remarkable feature of supercooled states
and glasses is the drastic increase in the
viscosity and the structural relaxation time that accompanies
non-exponentially observed in various time
correlation functions~\cite{Ediger1996Supercooled,
Ediger1996Supercooled, Jackle1999Models, Ngai2000Dynamic, Angell2000Relaxation,
Debenedetti2001Supercooled, Cavagna2009Supercooled, Ediger2012Perspective}.
Understanding the universal mechanism of the slow dynamics in glass transitions
is a challenging problem for condensed matter physics.

To tackle this problem, 
the notion of ``spatially heterogeneous dynamics'' or ``dynamic heterogeneity'' in 
glass-forming liquids has attracted much attention in recent decades
and has been considered central to understanding the slow
dynamics of glasses~\cite{Sillescu1999Heterogeneity, Ediger2000Spatially, Richert2002Heterogeneous,
Chandler2010Dynamics, Berthier2011Dynamic, Berthier2011Dynamical, Berthier2011Theoretical}.
Many theoretical, computational, and experimental efforts
have been devoted to understanding of dynamic heterogeneities in
glassy systems.

Experimentally, various nuclear magnetic
resonance (NMR) and other spectroscopic techniques have revealed 
``heterogeneous'' relaxation of the non-exponential 
decays in glassy systems,~\cite{Schmidt1991Nature,
Heuer1995Rate, Bohmer1996Dynamic, Bohmer1998Nature, Tracht1998Length, Russell2000Direct}.
In such systems, the non-exponential relaxation
can be explained as the superposition of individual particle contributions
with different relaxation rates~\cite{Ediger2000Spatially, Richert2002Heterogeneous}.

A large number of molecular simulations have also provided information
by allowing for the visualization of microscopic details regarding 
the molecular dynamics~\cite{Hurley1995Kinetic, Muranaka1995Beta, Kob1997Dynamical,
Yamamoto1997Kinetic, Donati1998Stringlike, Yamamoto1998Dynamics,
Yamamoto1998Heterogeneous, Perera1999Relaxation, Donati1999Growing,
Kim2000Apparent, Doliwa2000Cooperativety,
Glotzer2000Spatially, Lacevic2002Growing,
Lacevic2003Spatially, Berthier2004Time, Berthier2005Numerical, Pan2005Heterogeneity,
WidmerCooper2006Predicting, Brito2007Heterogeneous, Widmer2008Irreversible,
Cnadelier2010Spatiotemporal, Keys2011Excitations, Shiba2012Relationship}.
These simulations have demonstrated direct evidences of the dynamic heterogeneities,
\textit{i.e.}, molecular motions accompany correlated
domains and, to some extent, exceed the molecular scale in a
heterogeneous manner in both time and space.
Experiments to directly visualize these 
dynamic heterogeneities have also been performed in
colloidal glasses~\cite{Marcus1999Experimental,
Kegel2000Direct, Weeks2000Threedimentional, Weeks2002Properties,
Weeks2007Short, Prasad2007Confocal, Narumi2011Spatial, Hunter2012Physics}.

Those results have required 
characterizing and quantifying the length and time scales
to determine the physical role of dynamic heterogeneities in the underlying
mechanism of the glassy slow dynamics.
Furthermore, such information regarding spatio-temporal structures would
be indispensable to an assessment of the 
theoretical scenarios and hypotheses proposed thus far, such as the
Adam--Gibbs~\cite{Adam1965Temperature}, random
first-order transition~\cite{Kirkpatrick1989Scaling, Bouchaud2004On,
Lubchenko2007Theory}, dynamic
facilitation~\cite{Chandler2010Dynamics}, potential energy landscape~\cite{Heuer2008Exploring},
mode-coupling~\cite{Reichman2005Modecoupling, Gotze2009Complex},  
replicated liquid~\cite{Mezard1999Firstprinciple, Parisi2010Meanfield}, and
frustration-limited domain~\cite{Tarjus2005Frustrationbased} approaches.
Recent attention has focused on determining the physical origin
of dynamic heterogeneities by proposing 
various length scales including those governed by cooperative rearranging
regions~\cite{Karmakar2009Growing, 
Sengupta2012AdamGibbs},
mosaic length~\cite{Kirkpatrick1989Scaling, Xia2000Fragilities},
bond-orientational order (BOO)~\cite{Shintani2006Frustration, Kawasaki2007Correlation,
Kawasaki2010Structural, Tanaka2010Criticallike, Kawasaki2011Structural},
bond-breakage correlations~\cite{Yamamoto1997Kinetic,
Yamamoto1998Dynamics, Shiba2012Relationship},
icosahedral order~\cite{Dzugutov2002Decoupling, Pedersen2010Geometry, Leocmach2012Roles},
locally preferred structures (LPSs)~\cite{Coslovich2007Understanding,
Coslovich2011Locally}, geometrical frustration~\cite{Sausset2008Tuning, Sausset2010Growing,
Charbonneau2012Geometrical}, patch correlations~\cite{Kurchan2010Order,
Sausset2011Characterizing}, non-local
viscoelasticity~\cite{Furukawa2009Nonlocal, Furukawa2011Direct, Furukawa2012Dynamic},
Fickian diffusion~\cite{Berthier2004Time, Berthier2005Length, Szamel2006Time},
and point-to-set (PTS) correlations~\cite{Cavagna2007Mosaic, Biroli2008Thermodynamic, Kob2012Nonmonotonic,
Hocky2012Growing, Berthier2012Static, Karmakar2012Direct,
Cammarota2012Ideal}.

Recently, progress has been made in 
characterizing the dynamic length scale via multi-point correlations.
Extractions of the spatial
correlations between particle displacements during a typical time
interval can lead to the four-point correlation
function~\cite{Dasgupta1991Is, 
Franz2000On, Glotzer2000Timedependent, Glotzer2000Spatially, Donati2002Theory,
Lacevic2002Growing, Lacevic2003Spatially, Berthier2004Time, Whitelam2004Dynamic, 
Toninelli2005Dynamical, Chandler2006Lengthscale,
Szamel2006Four, Berthier2007Spontaneous, Berthier2007Spontaneous2}. 
Theoretical treatments have also been provided to analyze 
related multi-point susceptibilities based on the mode-coupling
approach~\cite{Biroli2004Diverging, Biroli2006Inhomogeneous,
Berthier2007Spontaneous, Berthier2007Spontaneous2, Szamel2010Diverging}.
Alternative 
multi-point correlations~\cite{Berthier2005Direct,
Berthier2007Spontaneous, Berthier2007Spontaneous2,
DalleFirrier2007Spatial, Brambilla2009Probing} and non-linear
susceptibilities~\cite{Bouchaud2005Nonlinear, Tarzia2010Anomalous,
CrausteThibierge2010Evidence, Diezemann2012Nonlinear} have also been proposed for
experimental measurements.

In addition, a deeper understanding of the time scale of heterogeneous
dynamics has been provided by 
non-linear responses such as those studies by multidimensional NMR, hole-burning
and photo-bleaching techniques~\cite{Schmidt1991Nature, Heuer1995Rate, Cicerone1995Relaxation,
Bohmer1996Dynamic, Wang1999How, Wang2000Lifetime}, and single molecule
measurements~\cite{Deschenes2002Heterogeneous,
Adhikari2007Heterogeneous, Zondervan2007Local, Mackowiak2009Spatial,
Mackowiak2011Probe, Bingemann2011Single}.
In these experiments, the lifetime of the dynamic
heterogeneity is evaluated from the exchange time between the slow- and
fast-moving regions, which is much
longer than the structural relaxation
time near the glass transition temperature.

In contrast, much
less attention has been paid to theoretical and computational
explorations of the characteristic time scale of
dynamic heterogeneities and its temperature dependence.
Note that a characterization of the time scale of dynamic heterogeneities
requires an analysis of the duration of the heterogeneous motions,
which essentially requires a multi-time extension of the four-point
correlation function.
Some studies have introduced relevant multi-time correlation functions
to investigate the time scale of heterogeneities~\cite{Heuer1997Heterogeneous, Heuer1997Information,
Doliwa1998Cage, Yamamoto1998Heterogeneous, Perera1999Relaxation, Qian2000Exchange,
VanZon2001Modecoupling, Doliwa2002How, Flenner2004Lifetime,
Jung2004Excitation, Jung2005Dynamical, Leonard2005Lifetime, 
Hedges2007Decoupling, Mizuno2010Lifetime, Mizuno2011Dynamical}.
However, several time intervals are fixed at a characteristic time scale,
and thus only limited information is available regarding the time scale
of the heterogeneous dynamics.

In our previous studies~\cite{Kim2009Multiple, Kim2010Multitime},
we have emphasized the importance of 
analyzing a four-point, three-time correlation function for various time intervals
to systematically quantify the temporal structures of dynamic
heterogeneities and their lifetimes.
From the three-time correlation function, we have found that the
lifetime increases and becomes much
longer than the $\alpha$-relaxation time $\tau_\alpha$ of the two-point
correlation function for a binary soft-sphere supercooled liquid.
The observed decoupling between the two time scales at low temperatures is
in agreement with the previously mentioned experimental results.
The exploited strategy is analogous to the multi-time correlations used in
recent multi-dimensional spectroscopy and other related optical techniques~\cite{Mukamel1999Principles,
Fayer2001Ultrafast, Khalil2003Coherent, Tanimura2006Stochastic,
Hochstrasser2007Twodimensional, Cho2009Two, Senning2010Subcellular,
Hamm2011Concepts, Berg2012Multidimensional}.
These spectroscopic techniques have become powerful
tools for examining the change from heterogeneous dynamics to 
homogeneous dynamics in various condensed phase systems~\cite{Asbury2004Water,Scimidt2005Pronounced,
Loparo2006Multidimensional2, Kraemer2008Temperature,
Paarmann2008Probing, GarrettRoe2008Threepoint, Yagasaki2008Ultrafast, Yagasaki2009Molecular}.
Furthermore, such multi-dimensional techniques have recently been applied to
supercooled and glassy states~\cite{Yagasaki2011Energy, Perakis2011TwoDimensional, King2012Ultrafast}.

The aim of the present paper is to investigate the spatio-temporal
structures of dynamic heterogeneities 
by numerically calculating the multi-point and multi-time correlation functions.
In particular, we extensively investigate the manner in which the
specifics of the model influence the extracted length and time scales.
Recently, the effects of the model details on the static
correlation and the dynamics have been critically examined for the
binary Kob--Andersen model and its Weeks--Chandler--Andersen
modification~\cite{Berthier2009Nonperturbative, Berthier2010Critical, Berthier2011The}.
While the pair correlations of these models are almost identical, the
structural relaxation times differ significantly.
These results necessitate an investigation of the 
model dependence on the length and time scales of the dynamic heterogeneities
extracted from multi-point and multi-time correlation functions.
Thus, in the present study, we employ a
frequently used binary mixture of Kob--Andersen, Wahnstr\"{o}m,
soft-sphere models, which 
exhibit a super-Arrhenius temperature dependence in their
structural relaxation times, and a network-forming strong liquid model
that exhibits Arrhenius behavior.
We provide comprehensive numerical results for the four-point correlation
function and the three-time correlation function extended from the four-point correlation.
Using our extensive numerical results,
we systematically determine the length and time scales of dynamic
heterogeneities for various glass models.

The paper is organized as follows.
In Sec.~\ref{model}, we introduce the simulation details of the glass-forming
models used in this paper.
In Sec.~\ref{results}, we present numerical calculations of the multi-point
and multi-time
correlation functions to characterize the spatio-temporal structures of 
the dynamic heterogeneities.
In Sec.~\ref{fskt}, we briefly
summarize numerical results using conventional two-point correlation
functions and determine the $\alpha$-relaxation time $\tau_\alpha$.
Then, in Sec.~\ref{four_point}, we evaluate the growing length
scales of the dynamics heterogeneities in the models by analyzing the
four-point correlation functions.
Finally, in Sec.~\ref{three_time}, the lifetimes of the dynamic
heterogeneities are quantified from the
three-time correlation functions, and  their dependence on the model and
fragility is examined.
In Sec.~\ref{summary}, we summarize our results and provide concluding remarks.

\section{simulations of model glasses}
\label{model}

In this study, we carry out extensive molecular dynamics simulations
for three-dimensional binary mixtures in the microcanonical ensemble.
The system contains $N_1$ particles of component 1 
and $N_2$ particles of component 2 under periodic boundary conditions.
The total number density is fixed at
$\rho=N/V$ with the total number of particles $N=N_1+N_2$ and a system volume $V$.

The models examined are
the well-known prototype models of glass-forming liquids: the
binary Kob--Andersen Lennard--Jones (KALJ) liquids~\cite{Kob1995Testing1, Kob1995Testing2}, the binary
Wahnstr\"{o}m (WAHN) liquids~\cite{Wahnstrom1991Moleculardynamics}, and
binary soft-sphere (SS) liquids~\cite{Bernu1985Molecular, Bernu1987Soft}.
In addition, we also study a model of a network-forming
(NTW) liquid that mimics $\mathrm{SiO_2}$ with short-range
spherical potentials~\cite{Coslovich2009Dynamics}.

\subsection{KALJ: binary mixture of Kob--Andersen Lennard--Jones particles}

The binary Lennard--Jones mixture is the most frequently utilized model
for the study of glass transitions~\cite{Kob1995Testing1, Kob1995Testing2}.
The pair potentials are given by
\begin{equation}
v_{\alpha\beta}(r)=4\epsilon_{\alpha\beta}\left[\left(\frac{\sigma_{\alpha\beta}}{r}\right)^{12}
-\left(\frac{\sigma_{\alpha\beta}}{r}\right)^{6}\right],
\label{eq_LJ}
\end{equation}
in which $\alpha, \beta  \in \{1, 2\}$ are the indexes of the particle species.
The energy and size ratios are $\epsilon_{12}/\epsilon_{11}=1.5$,
$\epsilon_{22}/\epsilon_{11}=0.5$ and 
$\sigma_{12}/\sigma_{11}=0.8$, $\sigma_{22}/\sigma_{11}=0.88$, respectively.
The masses of the two particle species are equal,  $m_1 = m_2 = 1$.
The interaction is truncated at $r = 2.5\sigma_{\alpha\beta}$.
The reduced
units $\sigma_{11}$, $\epsilon_{11}/k_B$, and 
$\sqrt{m_1\sigma_{11}^2/\epsilon_{11}}$ are used in the model for length,
temperature, and time, respectively.
The time step is $\Delta t=0.001$ in the reduced time units.
The total number density is fixed at $\rho=1.2$ with 
$N_1=800$ and $N_2=200$.
The temperatures investigated are $T=0.7, 0.65, 0.6, 0.55, 0.5$, and
$0.47$.

\subsection{WAHN: binary mixture of Wahnstr\"{o}m Lennard--Jones particles}

The KALJ model is a non-additive mixture and thus disobeys
the so-called Lorentz--Berthelot
combining rules due to the strong attraction between components 1 and 2.
Alternatively, the
 prototype model of the additive and equimolar binary Lennard--Jones mixture is
introduced by Wahnstr\"{o}m~\cite{Wahnstrom1991Moleculardynamics, Schroder2000Crossover}.
The interaction potentials are the same as in Eq.~(\ref{eq_LJ}), in
which 
the size, mass, and energy ratios are given as $\sigma_1 / \sigma_2 = 1/1.2$, $m_1 / m_2
= 1 / 2$, and $\epsilon_1 / \epsilon_2 = 1$, respectively.
The Lorentz--Berthelot rules,
\begin{align}
\sigma_{\alpha\beta} =\frac
{\sigma_\alpha + \sigma_\beta}{2}, &&
\epsilon_{\alpha\beta}
=\sqrt{\epsilon_\alpha  \epsilon_\beta},
\label{Lorentz_Berthelot}
\end{align}
are obeyed in this model.
Simulation results will be described in terms of the reduced 
units $\sigma_{1}$, $\epsilon_{1}/k_B$, and 
$\sqrt{m_1\sigma_{1}^2/\epsilon_{1}}$ for length,
temperature, and time, respectively.
The system consists of $N_1=500$ and 
$N_2=500$ particles with a fixed number density $\rho=0.75$.
A time step of $\Delta t=0.001$ is used.
The temperatures investigated are $T=0.8, 0.75, 0.7, 0.65, 0.6$, and
$0.58$.

\subsection{SS: binary mixture of soft-sphere particles}

We also study an equimolar binary mixture of soft-sphere
particles~\cite{Bernu1985Molecular, Bernu1987Soft}.
Particles interact via the soft-core potentials
\begin{equation}
v_{\alpha\beta}(r)=\epsilon_{\alpha\beta}\left(\frac{\sigma_{\alpha\beta}}{r}\right)^{12},
\label{eq_softcore}
\end{equation}
with 
the cubic smoothing function $v_{\alpha\beta}(r)=B(a-r)^3+C$ for 
distances $r > r_c=\sqrt{3}$.
The values of $a$, $B$, and $C$ are determined by the continuity
conditions up to the second derivative of $v_{\alpha\beta}(r)$.
The size, mass, and energy ratios are the same as those of the WAHN model:
$\sigma_1 / \sigma_2 = 1/1.2$, $m_1 / m_2= 1 / 2$, 
and $\epsilon_1 / \epsilon_2 = 1$, respectively.
Thus, this model can be regarded as a purely repulsive interacting system of
the WAHN model.
Simulation results will be described in terms of the reduced 
units $\sigma_{1}$, $\epsilon_{1}/k_B$, and
$\sqrt{m_1\sigma_{1}^2/\epsilon_{1}}$ for length,
temperature, and time, respectively.

The thermodynamic state of this model is usually characterized by the
following non-dimensional parameter:
\begin{equation}
\Gamma = \rho\left(\frac{\epsilon_1}{k_BT}\right)^{1/4} {l_0}^3,
\end{equation}
in which $l_0$ represents the effective particle size defined by $4{l_0}^3 =
(2\sigma_1)^3 + 2(\sigma_1+\sigma_2)^3 +(2\sigma_2)^3$.
In the simulation, 
the total number density is given as $\rho={l_0}^{-3}$ with $N_1=N_2=500$.
The investigated states are $\Gamma=1.30, 1.35, 1.38, 1.42, 1.45$, and $1.47$.
The corresponding temperatures are $T=0.350, 0.301, 0.276, 0.246,
0.226$, and $0.214$ with a time step of $\Delta t=0.005$.

\subsection{NTW: tetrahedral network-forming liquids}

In addition, we study a model of network-forming
liquids interacting via spherical short-ranged potentials~\cite{Coslovich2009Dynamics}.
This model is simple model and imitates $\mathrm{SiO_2}$ glasses,
in which tetrahedral networks strongly dominate the dynamics
with an Arrhenius behavior for the structural relaxation time, even near the
glass transition temperature.
The interaction potentials are given as
\begin{equation}
v_{\alpha\beta}(r)=\epsilon_{\alpha\beta}\left[\left(\frac{\sigma_{\alpha\beta}}{r}\right)^{12}
-(1-\delta_{\alpha\beta})\left(\frac{\sigma_{\alpha\beta}}{r}\right)^{6}\right],
\label{eq_ntw}
\end{equation}
in which $\delta_{\alpha\beta}$ is the Kronecker delta.
The interaction is truncated at $r = 2.5\sigma_{\alpha\beta}$.
The size and energy units are determined as follows:
\begin{align}
\sigma_{12} / \sigma_{11} &= 0.49 & \sigma_{22} / \sigma_{11} &= 0.85\\
\epsilon_{12} / \epsilon_{11} &= 24 & \epsilon_{22} / \epsilon_{11} &= 1.
\end{align}
The mass ratio is determined as $m_2 / m_1 = 0.57$ from the same ratio of
O and Si.
The units of length, time, and temperature are given as 
$\sigma_{11}$, $\epsilon_{11}/k_B$, and 
$\sqrt{m_1\sigma_{11}^2/\epsilon_{11}}$, respectively.
These parameters are adjusted to reproduce the radial distribution
functions of the $\mathrm{SiO_2}$ amorphous states.
Tetrahedral networks are found to be formed
due to the highly asymmetric size ratio and the strong attraction
between the different components~\cite{Coslovich2009Dynamics}.
The density of the investigated system is $\rho=1.655$, with particle numbers
$N_1=1,000$ and $N_2=2,000$.
This value corresponds to the density $\rho = 2.37 \mathrm{g \mathring{A}^{-3}}$ of the
so-called van Beest--Kramer--van Santen (BKS) model for the silica
glass~\cite{vanBeest1990Force, Horbach1996Finite, Horbach1999Static}.
The simulations are carried out at $T=0.42, 0.4, 0.38, 0.36, 0.34$ and
$0.32$ with a time step of $\Delta t = 0.0005$.

\section{Results and Discussion}
\label{results}
\subsection{Intermediate scattering function and the $\alpha$-relaxation time}
\label{fskt}

\begin{figure}[t]
\includegraphics[width=.48\textwidth]{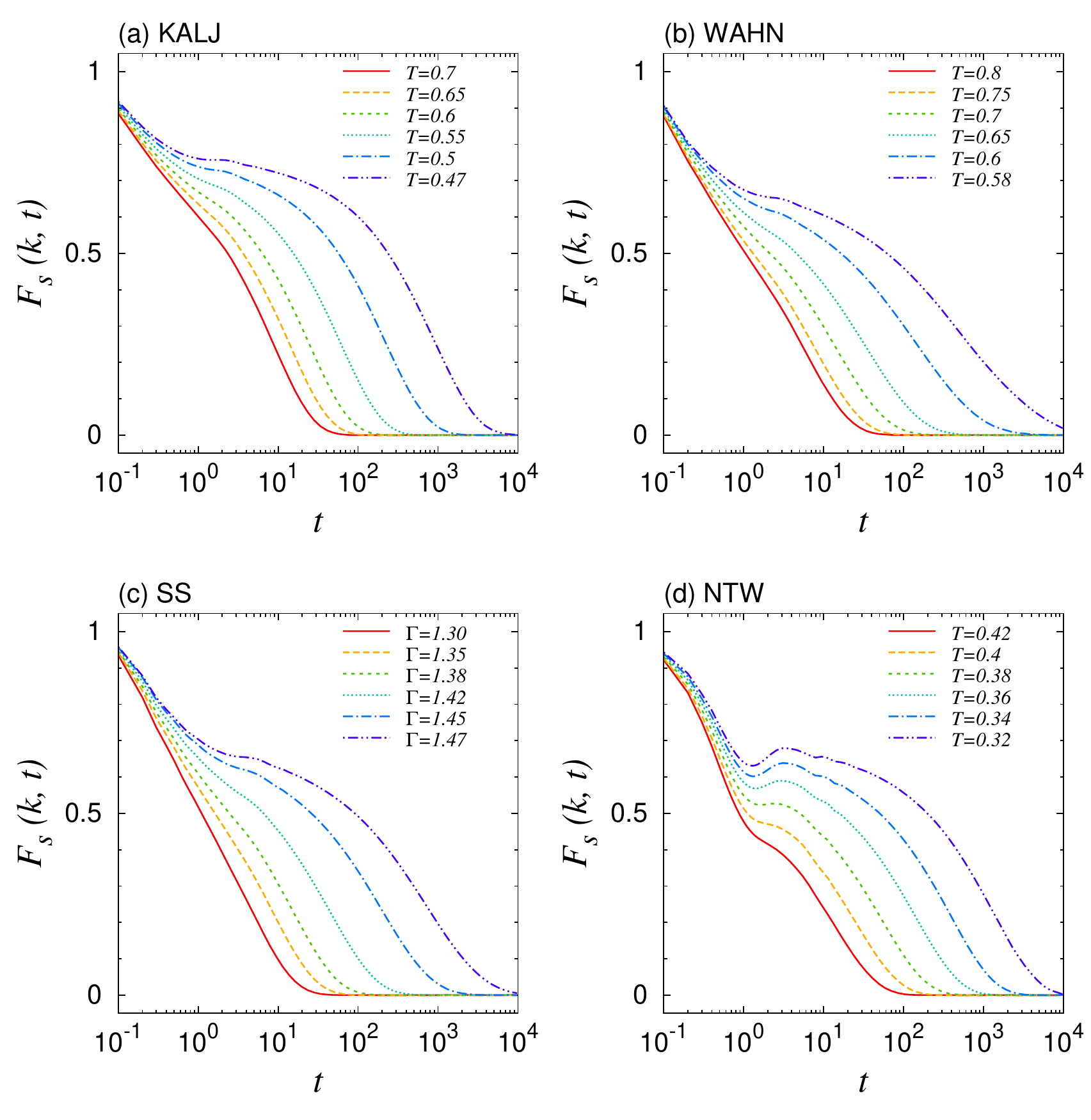}
\caption{
Self-part of the intermediate scattering function $F_s(k, t)$ of the
 component 1 particles for various glass-forming liquid models: (a)
 KALJ, (b) WAHN, (c) SS, and (d) NTW.
}
\label{fig_fskt}
\end{figure}
\begin{figure}[t]
 \includegraphics[width=.48\textwidth]{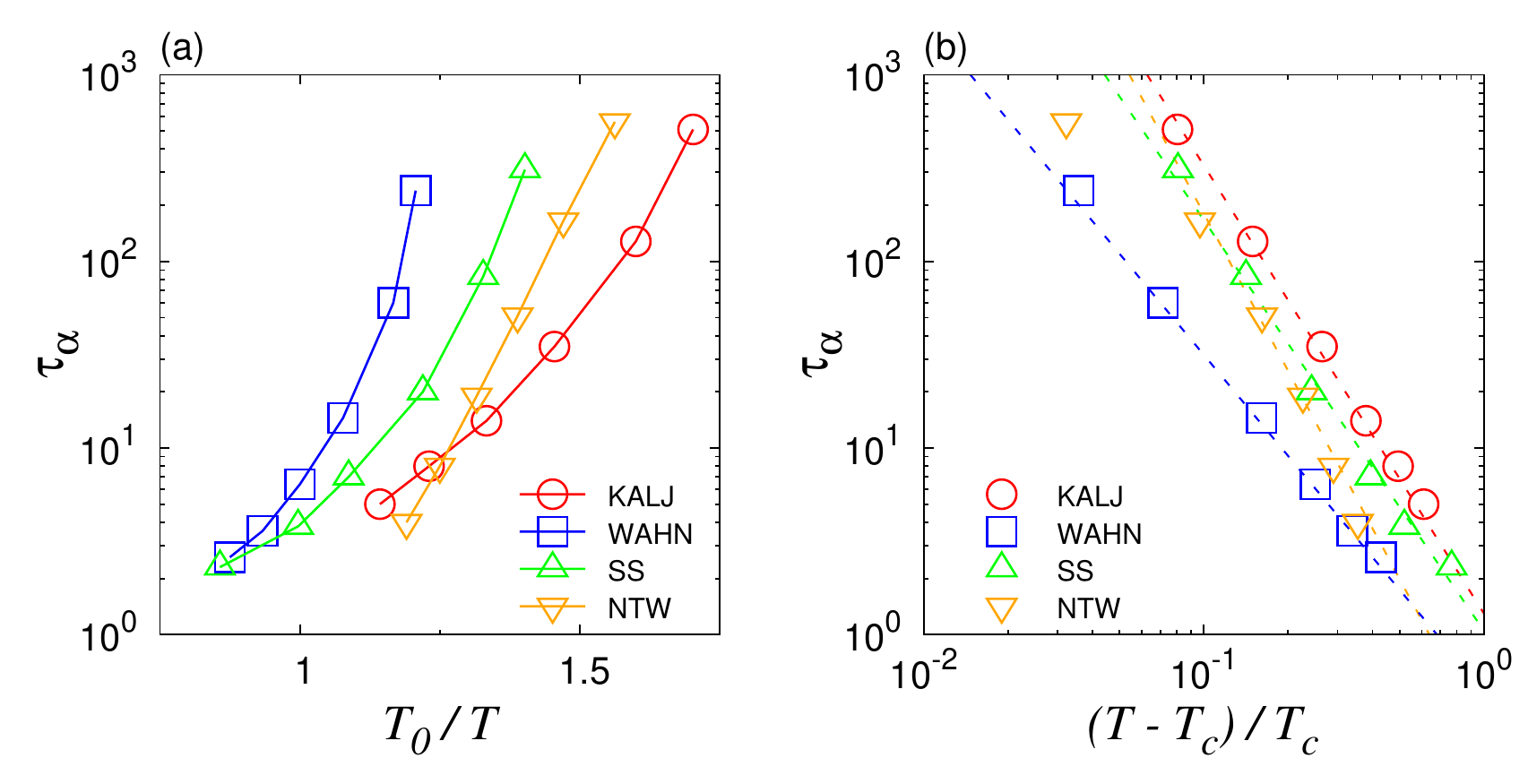}
\caption{
(a) $\alpha$-relaxation time $\tau_\alpha$ as a function of the inverse
 temperature $T_0/T$ with the onset temperature $T_0$.
(b) $\alpha$-relaxation time $\tau_\alpha$ as a function of the
 temperature difference $(T-T_c)/T_c$ from the mode-coupling 
 divergence temperature $T_c$. Each dashed line represents the power-law fits
 with exponent $\Delta=2.4$, $1.8$, $2.2$, and $2.8$ for the KALJ, WAHN, SS,
 and NTW models, respectively.
}
\label{fig_tau}
\end{figure}

First,
we study the conventional two-point density correlation function and
determine the structural $\alpha$-relaxation time $\tau_\alpha$.
The self-part of the intermediate scattering
function of the component $1$ particles
\begin{equation}
F_s(k, t)=\left\langle \frac{1}{N_1}\sum_{j=1}^{N_1}
\exp[i\bm{k}\cdot\Delta \bm{r}_j(0, t)]\right\rangle,
\label{eq_fskt}
\end{equation}
is calculated for various glass-forming models.
Here, $\Delta \bm{r}_j(0, t)\equiv
\bm{r}_j(t)-\bm{r}_j(0)$ is the $j$th particle displacement vector at 
times $0$ and $t$, and $\bm{k}$ is the wave vector.

The behavior of $F_s(k, t)$ at various temperatures is demonstrated 
in Fig.~\ref{fig_fskt}.
The wave vector $k=|\bm{k}|$ is chosen so that the static
structure factor of component 1, $S_{11}(k)$, marks the main peak as (a) $k=7.25$, (b)
$6.65$, (c) $6.55$, and (d) $8.0$ for the KALJ, WAHN, SS, and NTW models,
respectively.
From these calculations, we determine the $\alpha$-relaxation time $\tau_\alpha$ as
$F_s(k, \tau_\alpha)=e^{-1}$.
In Fig.~\ref{fig_tau}(a), the temperature
dependence of $\tau_\alpha$ is plotted as a function of the inverse temperature $T_0/T$.
Here $T_0$ denotes the onset temperature introduced in the earlier work~\cite{Berthier2009Nonperturbative}.
We set $T_0$ as $T_0=0.8$, $0.7$, $0.3$, and $0.5$ for the KALJ, WAHN, SS,
and NTW models, respectively.
Below this onset temperature, $F_s(k, t)$ begins to develop a two-step
relaxation in each model.
Furthermore, the temperature dependence of $\tau_\alpha$ exhibits
super-Arrhenius behavior in the KALJ, WAHN, and SS models.
This behavior is typical for the fragile glass-forming liquids.
In contrast, in the NTW model the tetrahedral networks begin to develop
strongly below $T_0$.
Correspondingly, the temperature dependence of the $\alpha$-relaxation time obeys
the Arrhenius law~\cite{Coslovich2009Dynamics}.
In addition, we summarize the power-law behavior as $\tau_\alpha \propto |T-T_c|^{-\Delta}$,
as predicted by the mode-coupling theory~\cite{Reichman2005Modecoupling, Gotze2009Complex}.
The results of the power-law fittings for the four model liquids are
demonstrated in Fig.~\ref{fig_tau}(b).
We obtain the values of the exponent $\Delta$ and the mode-coupling temperature
$T_c$ as $(\Delta, T_c) \approx (2.4, 0.435), (1.8, 0.56), (2.2,
0.198)$, and $(2.8, 0.31)$ for KALJ, WAHN, SS, and NTW liquids,
respectively.
Similar results have been reported for the 
KALJ, SS, and NTW models~\cite{Berthier2012Finitesize}.
We also note that the power-law behavior of the NTW model is reliable over the limited
temperature range, as investigated in the simulations for the BKS
model~\cite{Horbach1999Static}.

\subsection{Four-point correlations and the growing length scale of dynamic heterogeneities}
\label{four_point}

\begin{figure}[t]
\includegraphics[width=.48\textwidth]{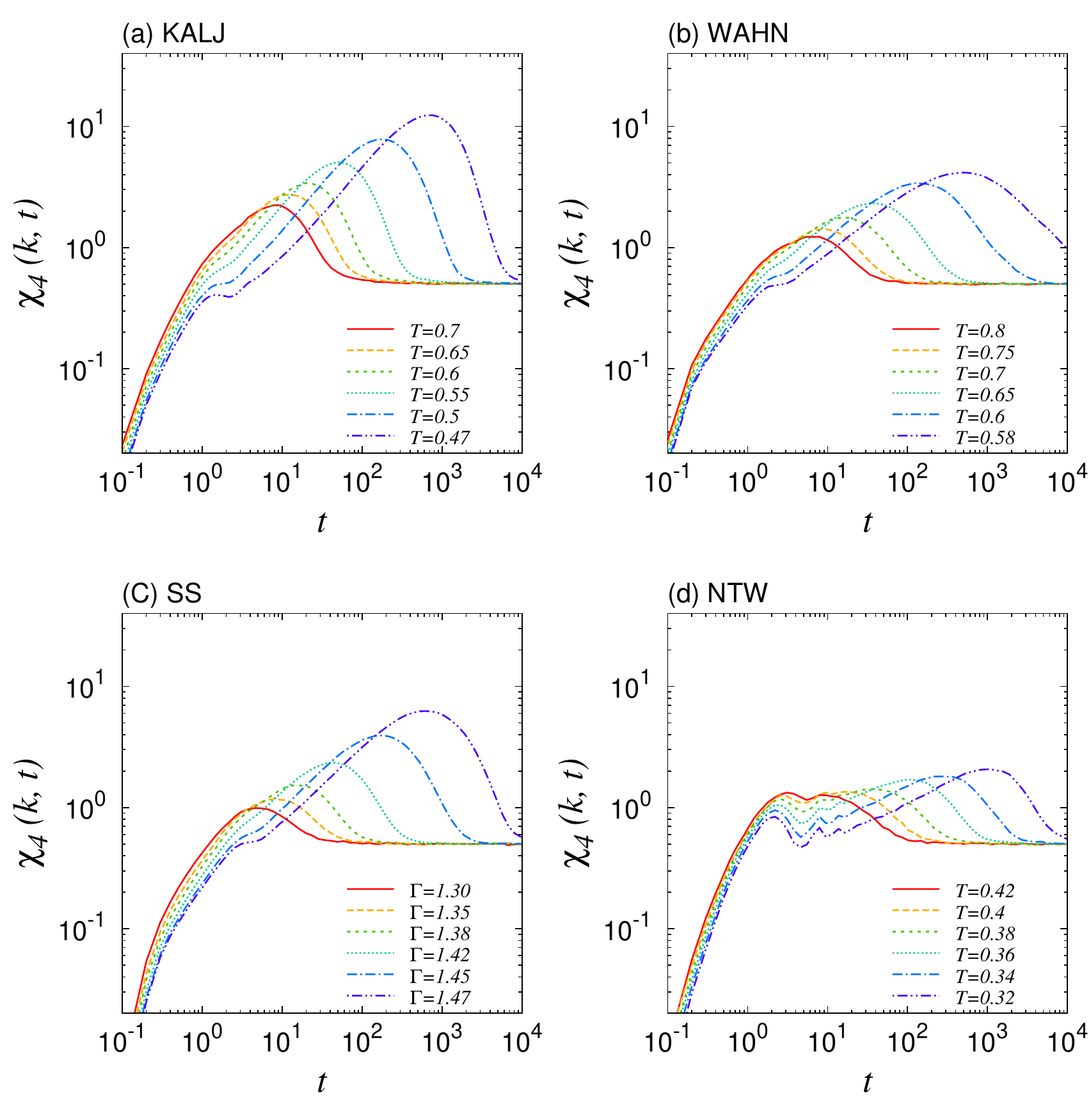}
\caption{
Four-point dynamical susceptibility $\chi_4(k, t)$ for the (a) KALJ, (b)
 WAHN, (c) SS, and (d) NTW models.
}
\label{fig_chi4}
\end{figure}

To characterize the growth of the dynamic heterogeneities in supercooled
states, 
the four-point correlation function is introduced to measure
the correlation of the particle mobility field at a given time interval.
There are several choices for the mobility field such as the
the particle displacement amplitude~\cite{Donati1999Growing},
the overlap function~\cite{Lacevic2003Spatially}, and
the intermediate scattering function~\cite{Berthier2004Time}.
Physically, the choice of the mobility field does not alter the
fundamental meaning of the four-point correlation function.
We here calculate
the four-point dynamical susceptibility $\chi_4(k, t)$, which is defined as the
variance of the
self-part of the intermediate scattering function $F_s(k, t)$:
\begin{equation}
\chi_4(k, t) = N_1 [ \langle \hat F_s(\bm{k}, t)^2 \rangle- \langle
\hat{F_s}(\bm{k}, t)\rangle ^2 ].
\label{chi4_eq2}
\end{equation}
We utilize $\hat F_s(\bm{k}, t)$ expressed as
\begin{equation}
\hat F_s(\bm{k}, t)=
\frac{1}{N_1}\sum_{j=1}^{N_1} \cos[\bm{k}\cdot\Delta \bm{r}_j(0, t)],
\end{equation}
with $F_s(k, t)=\langle \hat{F}_s(\bm{k}, t) \rangle$.
The total value of $\chi_4(k, t)$ is approximately proportional to the
extension of the
spatial correlations in dynamics with $k$ at a given time interval $t$
because $\chi_4(k, t)$ investigates the increasing deviation 
of the two-point correlation function $F_s(k, t)$ from the mean behavior,
As is well documented in various
studies~\cite{Toninelli2005Dynamical, Szamel2006Four, Chandler2006Lengthscale, Berthier2007Spontaneous} and as
demonstrated in Fig.~\ref{fig_chi4},
$\chi_4(k, t)$ typically has its maximum value at the time scale
of $\tau_\alpha$, which increases as the temperature decreases.
The growth of $\chi_4(k, t)$ for the strong NTW liquid is
suppressed and is smaller than those of other fragile liquids.
This behavior implies that the dynamic heterogeneity is less pronounced
and plays a minor role 
in the strong liquid, as revealed in previous
studies~\cite{Vogel2004Temperature, Berthier2007Revisiting, Berthier2007Spontaneous}.
Therefore, 
the dynamics of the strong liquid can be interpreted as mostly occurring
through the rearrangement of strongly connecting tetrahedral networks.
Interestingly, a similar suppression of $\chi_4(k, t)$ in strong liquids, for longer times, has been
observed in polydispersed systems~\cite{Abraham2008Suppression,
Kawasaki2011Structural}, colloidal gels~\cite{Abete2007Static, Fierro2008Dynamical},
and confined systems in random media~\cite{Kim2009Slow, Kim2011Slow}.

\begin{figure}[t]
\includegraphics[width=.48\textwidth]{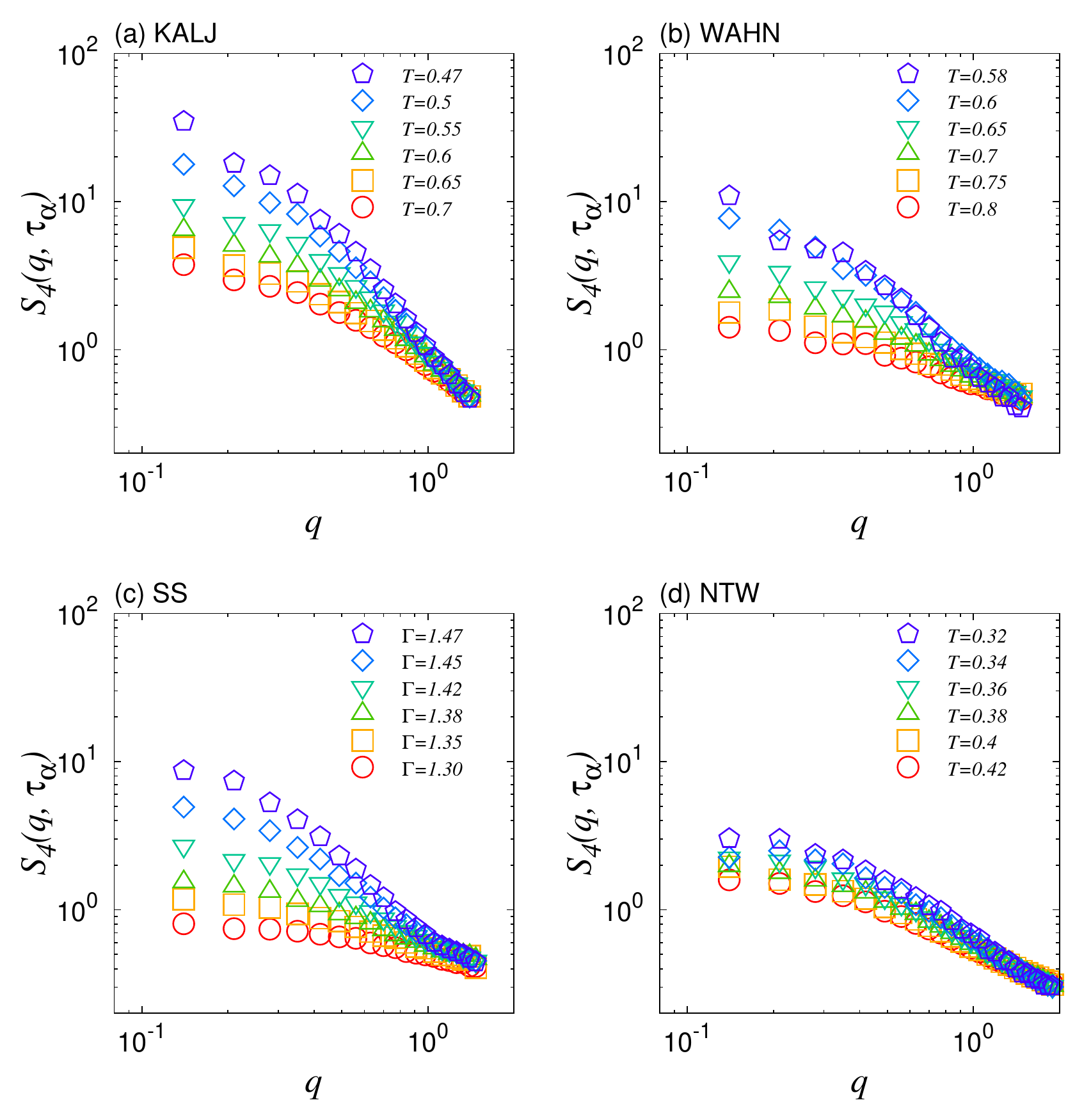}
\caption{
Four-point static structure factor $S_4(q, \tau_\alpha)$ at various
 temperatures as a
 function of the wave-number $q$ for the (a) KALJ, (b)
 WAHN, (c) SS, and (d) NTW models.
}
\label{fig_s4qt}
\end{figure}
\begin{figure}[t]
\includegraphics[width=.48\textwidth]{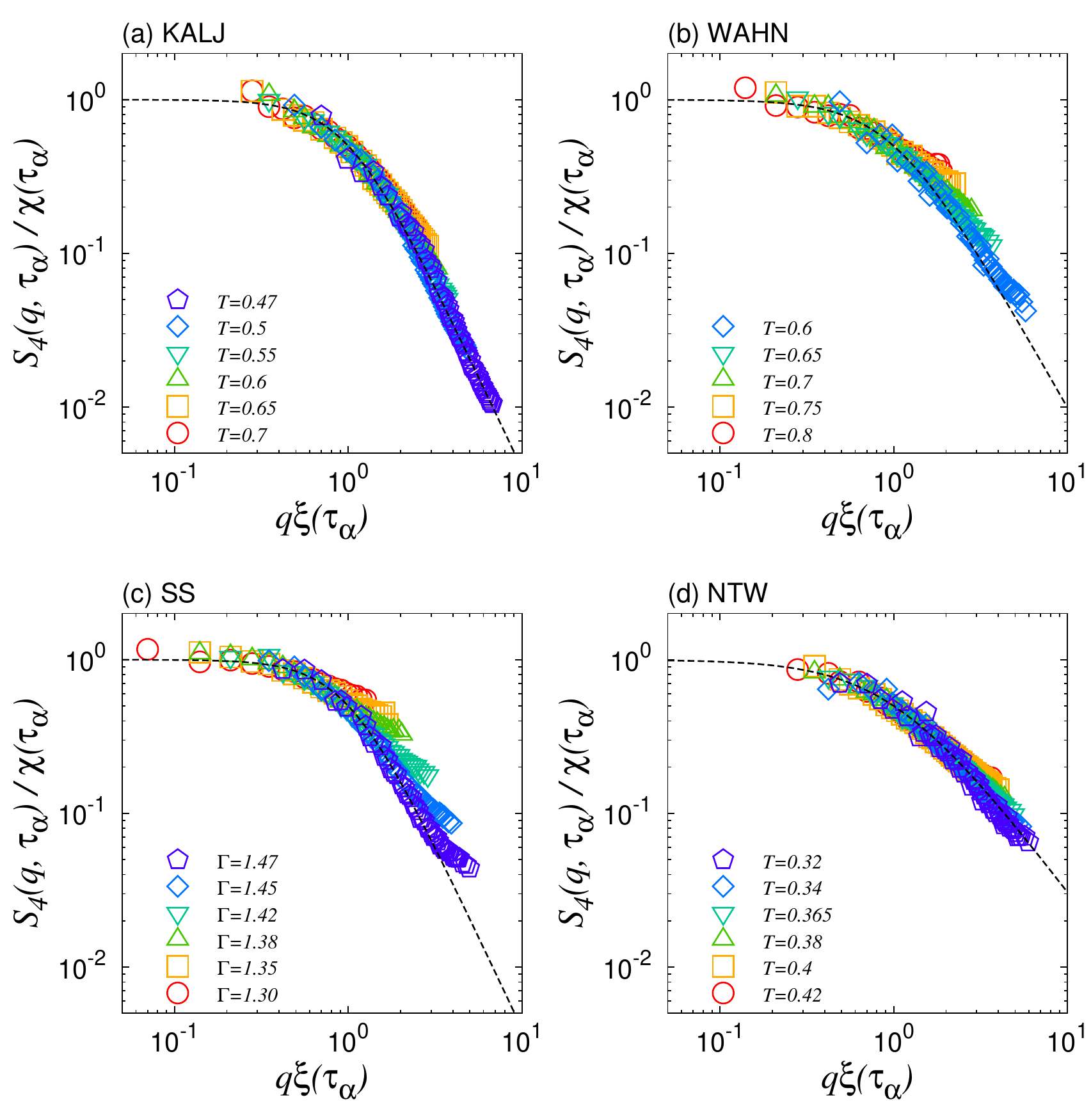}
\caption{
Scaled four-point static structure factor $S_4(q, \tau_\alpha) / \chi(\tau_\alpha)$
as a function of $q\xi(\tau_\alpha)$ at various temperatures
for the (a) KALJ, (b) WAHN, (c) SS, and (d) NTW models.
The dashed line represents the Ornstein--Zernike form
$1/(1+(q\xi(\tau_\alpha))^\alpha)$ with (a) $\alpha=2.4$, (b) $2.0$, (c)
 $2.4$, and (d) $1.5$.
}
\label{fig_s4qt_scaling}
\end{figure}

The study of the spatial correlation of the
four-point correlation function is also essential
to extract the growing length scales $\xi$ of the dynamic heterogeneities.
In this study, instead of the self-intermediate scattering function, we
utilize the frequently used four-point correlation function
using the overlap function, which is defined as follows:
\begin{equation}
S_4(q, t) = \frac{1}{N_1}\langle Q(\bm{q}, t) Q(-\bm{q}, t)\rangle,
\label{eq_s4qt}
\end{equation}
\begin{equation}
Q(\bm{q}, t) =\sum_{j=1}^{N_1} W_j(a, t) \exp[-i \bm{q}\cdot \bm{r}_j(0)],
\end{equation}
with the wave vector $q = |\bm{q}|$.
Here, $W_j(a, t) = \Theta(a - |\bm{r}_j(t) - \bm{r}_j(0)|)$ is the
overlapping function with the Heaviside step's function
$\Theta(x)$~\cite{Glotzer2000Spatially, Lacevic2002Growing, Lacevic2003Spatially}.
The function $W_j(a, t)$ selects a particle that moves further than the
distance $a$ during the time interval $t$.
The value $a=0.3$ is typically chosen.
As studied in the previous study~\cite{Lacevic2003Spatially},
the relaxation profile of the overlap function $Q(t)=
(1/N_1)\langle\sum_{j=1}^{N_1}W_j(a, t) \rangle$ with this probe length
scale $a=0.3$ approximately corresponds to that of the
self-intermediate scattering function $F_s(k, t)$ with the wave number
marking the main peak of the static structure factor.
We also note that any significant differences in $S_4(q, t)$ are not
observed if we choose $F_s(k, t)$ as the mobility field.
Although recent reports have described the results of $S_4(q, t)$ using
large-scale simulations~\cite{Stein2008Scaling, Karmakar2010Analysis,
Flenner2010Dynamic, Flenner2011Analysis, Mizuno2011Dynamical}, we
revisit the identification of the correlation length $\xi$ from the four-point correlation
function, Eq.~(\ref{eq_s4qt}).
Accordingly, we use much larger systems with $N=100,000$ for the KALJ, WAHN
and SS liquids and $N=90,000$ for the NTW liquid to calculate
$S_4(q, t)$.
The other parameters described in Sec.~\ref{model} are unchanged throughout the simulations.
Correspondingly, the linear dimension of the system, in termed of the
unit length, $L=V^{1/3}$, is given as $L=43.68$, $51.09$, $51.27$, and $37.89$
for the KALJ, WAHN, SS, and NTW models, respectively.

\begin{figure}[t]
\includegraphics[width=.48\textwidth]{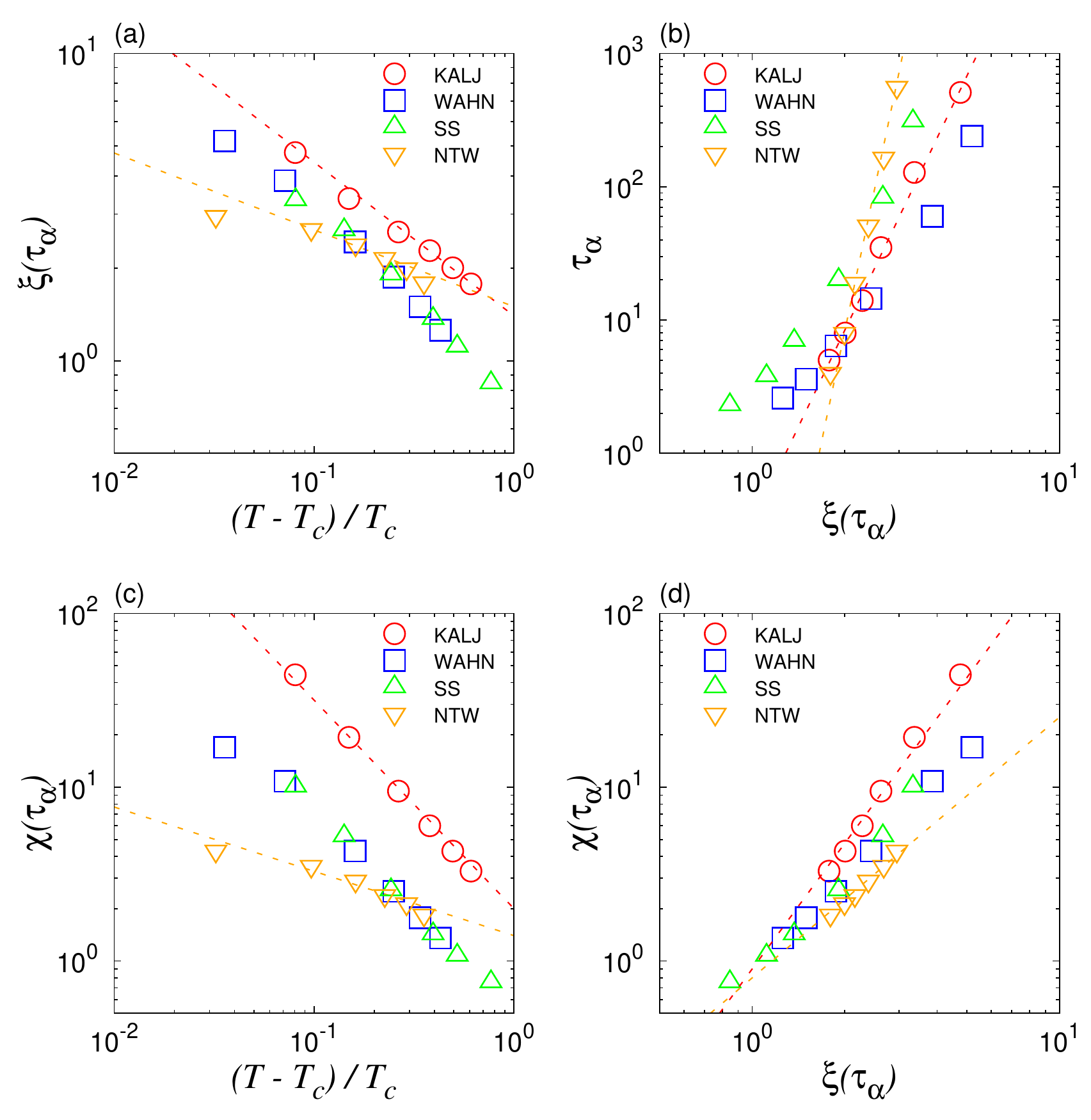}
\caption{
(a) Correlation length $\xi(\tau_\alpha)$ as a function of $(T-T_c)/T_c$.
The dashed line represents the power-law behavior, $\xi(\tau_\alpha) \sim
 |T-T_c|^{-\nu}$ with $\nu=0.5$ (red) and $\nu=0.25$ (orange),
 respectively.
(b) The relationship between the correlation length $\xi(\tau_\alpha)$ and
the $\alpha$-relaxation time $\tau_\alpha$.
The dashed line represents the
 power-law behavior, $\tau_\alpha \sim \xi(\tau_\alpha)^{z}$ with
 $z=4.4$ (red) and $z=11.0$ (orange), respectively.
(c) The dynamic susceptibility $\chi(\tau_\alpha)$ as a function of $(T-T_c)/T_c$.
The dashed line represents the power-law behavior, $\chi(\tau_\alpha) \sim
 |T-T_c|^{-\gamma}$ with $\nu=1.2$ (red) and $\nu=0.37$ (orange),
 respectively.
(d) The relationship between the correlation length $\xi(\tau_\alpha)$ and
the dynamic susceptibility $\chi(\tau_\alpha)$.
The dashed line represents the 
 power-law behavior, $\chi(\tau_\alpha) \sim \xi(\tau_\alpha)^{2-\eta}$ with
 $2-\eta=2.4$ (red) and $2-\eta=1.5$ (orange), respectively.
}
\label{fig_xi4}
\end{figure}

Figure~\ref{fig_s4qt} shows the wave-number dependence of $S_4(q, t)$ on
the time scale $\tau_\alpha$.
As indicated in Fig.~\ref{fig_s4qt},
$S_4(q, t)$ at the $\alpha$-relaxation time grows in small wave-numbers of $q$,
particularly at low temperatures.
This observation indicates that the mobile (or immobile) particles become
highly correlated in space when the system undergoes supercooling.
The behavior of $S_4(q, t)$ at small wave-numbers can be described by
the Ornstein--Zernike (OZ) form:
\begin{equation}
S_4(q, t) = \frac{\chi(t)}{1 + (q\xi(t))^{\alpha}},
\label{sk4_goz}
\end{equation}
in which $\xi(t)$ is regarded as the length scale of the dynamic heterogeneity and $\chi(t)$ is the
dynamic susceptibility at $q\to 0$.
The OZ form with $\alpha=2$ has been
used in previous studies~\cite{Yamamoto1998Dynamics, Lacevic2003Spatially, Berthier2004Time}.
However, Fig.~\ref{fig_s4qt} shows that the exponent $\alpha$ depends on
the details of the simulation model.
Figure~\ref{fig_s4qt_scaling} displays the scaled function $S_4(q,
\tau_\alpha)/\chi(\tau_\alpha)$ as a function of $q\xi(\tau_\alpha)$.
The results are in good agreement with $\alpha \approx 2.4$, $2.0$,
$2.4$, and $1.5$ for the KALJ, WAHN, SS, and NTW models, respectively.
The apparent difference in the exponent
$\alpha$ between the fragile (KALJ, WAHN, and SS) and strong (NTW)
glass-forming liquids may
be related to the change in geometrical characteristics of the heterogeneous
motions if we employ an analogy to the critical phenomena~\cite{Onuki2007Phase}.
A similar difference in the exponent has been found in kinetically constrained
models (KCMs), in which a snapshot of the dynamic heterogeneity in
strong KCM model
exhibits a smoother cluster structure than in fragile KCM model~\cite{Pan2005Heterogeneity}.

The values of the length scales $\xi$
and the susceptibility $\chi$ at the
$\alpha$-relaxation time are determined from the fitting to Eq.~(\ref{sk4_goz}).
Figure~\ref{fig_xi4}(a) shows the temperature dependence of the
correlation length $\xi(\tau_\alpha)$.
We find that the increasing length scale $\xi(\tau_\alpha)$ with decreasing
temperature can be described by the mode-coupling-like power-law
$\xi(\tau_\alpha) \sim |T-T_c|^{-\nu}$ at the investigated temperatures.
The exponent is $\nu\approx 0.5$ for the fragile KALJ, WAHN, and
SS models, whereas $\nu$ decreases to $\nu \approx 0.25$ for the
strong NTW model.
However, the length $\xi$ at lowest temperature $T=0.32$ is apparently
deviated from the power-law behavior of the NTW model, which is a strong
liquid exhibiting the Arrhenius behavior.
This limited power-law behavior is also observed in the temperature
dependence of the $\alpha$-relaxation time (see Fig.~\ref{fig_tau}(b)).
In addition, we obtained the scaling relationship $\tau_\alpha \sim
\xi(\tau_\alpha)^z$ found in Fig.~\ref{fig_xi4}(b).
In analogy to the dynamical critical phenomena, 
the exponent $z$ is approximately determined by the
relationship $z=\Delta/\nu$, \textit{i.e.}, the exponent $z$ becomes
$z\approx 4.4$, $4.8$, $3.6$, and $11.0$ for the KALJ, WAHN, SS, and NTW
models, respectively.

We also examine the relationship between the susceptibility
$\chi(\tau_\alpha)$ and the correlation length $\xi(\tau_\alpha)$ via
the scaling $\chi(\tau_\alpha) \sim \xi(\tau_\alpha)^{2-\eta}$,
as demonstrated in Figure~\ref{fig_xi4}(d) 
These results indicate that the scaling exponent of each model is correlated to the
value of the exponent $\alpha$ utilized in the OZ function, as shown in Eq.~(\ref{sk4_goz}).
In addition, the dynamic
susceptibility $\chi(\tau_\alpha)$ is approximated by the scaling
relation $\chi(\tau_\alpha)\sim |T-T_c|^{-\gamma}$ with
$\gamma=(2-\eta)\nu$, as shown in Fig.~\ref{fig_xi4}(c).
Similar dynamic criticality has been investigated in
Refs.~\onlinecite{Whitelam2004Dynamic, Pan2005Heterogeneity}.

Here, we remark that a recent theoretical treatment has been presented based on 
mode-coupling approach, referred to as the inhomogeneous
mode-coupling theory (IMCT)~\cite{Biroli2006Inhomogeneous, Szamel2010Diverging}.
The IMCT predicts the scaling exponents to be $\nu=1/4$ and
$2-\eta=4$ at the $\alpha$-relaxation time scale.
To examine the validity of these predictions, intensive large-scale molecular simulations
have been carried out~\cite{Stein2008Scaling,
Karmakar2010Analysis, Flenner2010Dynamic, Flenner2011Analysis}.
Our exponents, $z \approx 4.4$ and $2-\eta \approx 2.4$, of the KALJ model
are similar to the values obtained in
Refs.~\onlinecite{Berthier2007Spontaneous, Karmakar2010Analysis}.
In these findings, the
molecular simulations disagree with the theoretical predictions.
However, the IMCT does
\textit{not} investigate a four-point correlation function such as
Eq.~(\ref{eq_s4qt}) but treats
the three-point correlation defined as the response function of the
two-point correlation function due to the inhomogeneous external field.
Further analysis is needed to
understand the relationship between the four-point correlations utilized here and the
inhomogeneous three-point susceptibilities.

\subsection{Three-time correlations and lifetimes of dynamic heterogeneities}
\label{three_time}

\begin{figure}[t]
\includegraphics[width=.48\textwidth]{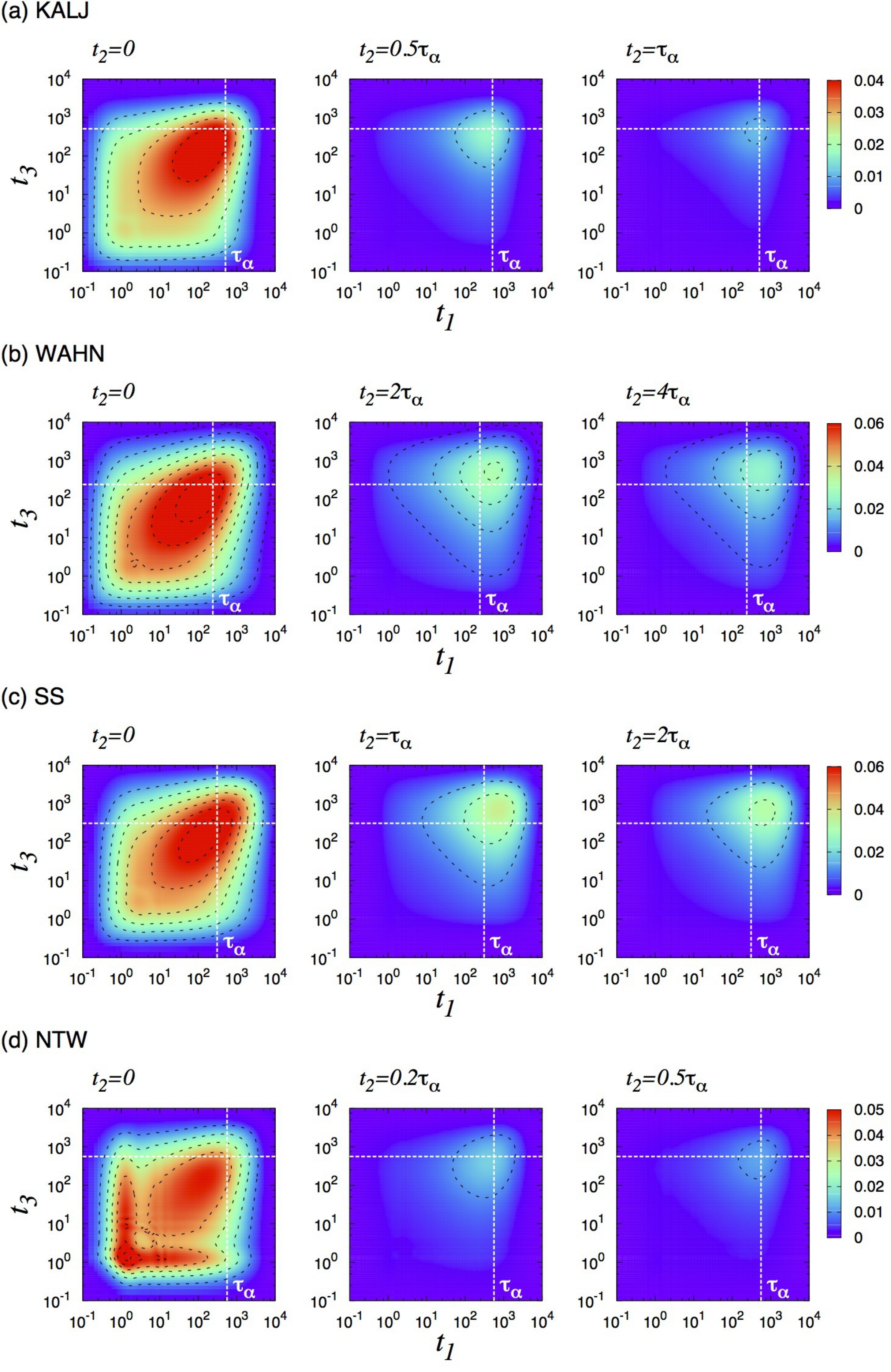}
\caption{
Two-dimensional correlation maps of the three-time correlation
 functions $\Delta F_4(k, t_1, t_2, t_3)$
for the (a) KALJ, (b) WAHN, (c) SS, and (d) NTW models.
The state is chosen at the lowest temperature for each model as (a) $T=0.47$, (b)
 $T=0.58$, (c) $\Gamma=1.47$, and (d) $T=0.32$.
The waiting times $t_2$ normalized by $\tau_\alpha$ are increased from left to right
in each model.
The vertical (horizontal) dashed line denotes the $\alpha$-relaxation time as
 $t_1=\tau_\alpha$ ($t_3=\tau_\alpha$).
Note that the waiting time is different in each figure.
}
\label{fig_F4}
\end{figure}
\begin{figure}[t]
\includegraphics[width=.48\textwidth]{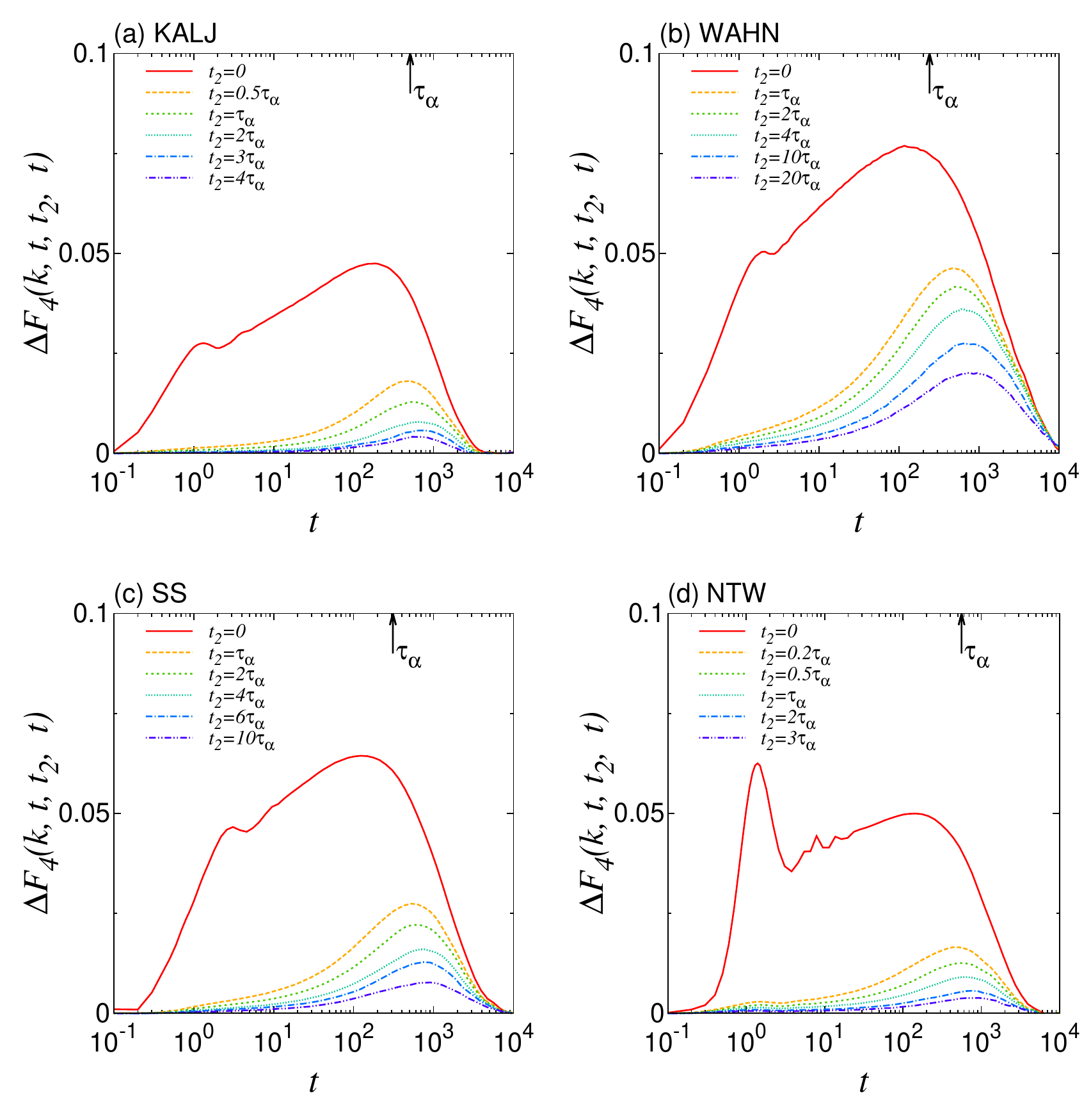}
\caption{
The diagonal portions of the three-time correlation functions $\Delta F_4(k, t, t_2, t)$
for the (a) KALJ, (b) WAHN, (c) SS, and (d) NTW models.
The state is chosen at the lowest temperature for each model as (a) $T=0.47$, (b)
 $T=0.58$, (c) $\Gamma=1.47$, and (d) $T=0.32$.
The waiting times $t_2$ are normalized by $\tau_\alpha$ at each temperature.
}
\label{fig_F4_t1t3}
\end{figure}

In this subsection,
we explore and accentuate the characteristic time scale of the dynamic heterogeneity.
To this end, we search for the multi-time extension of the four-point correlation function
$\chi_4(k, t)$ by introducing
\begin{equation}
\Delta F_4(k, t_1, t_2, t_3)
 = \biggl\langle  
\frac{1}{N_1}\sum_{j=1}^{N_1}  \delta F_j(\bm{k}, \tau_2, \tau_3)
\delta F_j(\bm{k}, 0, \tau_1))\biggr\rangle,
\label{three_time_correlation}
\end{equation}
in which
\begin{equation}
\delta F_j(\bm{k}, 0, t) = \cos[\bm{k}\cdot \Delta \bm{r}_j(0, t)] -
 F_s(k, t)
\end{equation}
provides the individual fluctuations in the two-point correlation function
at times $0$ and $t$.
This three-time correlation function examines the correlations at
four different times, $0$, $\tau_1$, $\tau_2$, and $\tau_3$.
In practice, $\Delta F_4(k, t_1, t_2, t_3)$ reveals the correlation of
the two-point correlation function $F_s(k, t)$ between the two time
intervals, $t_1=\tau_1$ and $t_3=\tau_3-\tau_2$.
Furthermore, 
the progressive changes in the waiting time
$t_2=\tau_2-\tau_1$ of the
three-time correlation function
$\Delta F_4(k, t_1, t_2, t_3)$
allow for an investigation of how the correlated motions
change with time $t_2$.
In fact, the first time-interval portion $\delta F_j(\bm{k}, 0, \tau_1)$
enables a distinction between the subensemble of slow- and fast- moving particles
in the dynamic heterogeneities.
In addition, the total function $\delta F_j(\bm{k}, 0, \tau_1)\delta
F_j(\bm{k}, \tau_2, \tau_3)$ can reveal how long
the correlations in the dynamics of the subensembles remain over the waiting
time $t_2$~\cite{Kim2009Multiple, Kim2010Multitime}.
Therefore, this multi-time correlation function can provide the temporal
structures and the associated characteristic time scale of the dynamic heterogeneity.
In other words, the time scale extracted from the three-time
correlation function can be regarded as the lifetime of the dynamic heterogeneity, which
should be associated with the length scale $\xi$ determined from the
four-point correlation function.

Figure~\ref{fig_F4} shows the time evolutions of the three-time
correlation functions $\Delta F_4$ at the lowest temperature for each
glass-forming liquid.
Note that the wave-number $k$ is chosen as the same value used in the
calculation of $F_s(k, t)$ in Fig.~\ref{fig_fskt}.
We also provide the
diagonal portions of the time evolutions along $t=t_1=t_3$ for various $t_2$ in
Fig.~\ref{fig_F4_t1t3}.
In previous studies~\cite{Kim2009Multiple, Kim2010Multitime},
we have examined the three-time correlation functions for the SS model.
This work confirms that the basic features of $\Delta F_4$ are similar and
that the time evolutions at $t_2$ occur similarly among all of the glass
models studied: 
(i) As the temperature decreases, 
the intensity of the three-time correlation $\Delta F_4(k, t_1, t_2,
t_3)$ gradually increases (see the results for other temperatures in
the Supplementary Material~\cite{Kim2012Supplement}).
This result is due to the correlations between particles that move 
slower (faster)  during the first time interval $t_1$ and
remain slower (faster) during the second time interval $t_3$.
(ii) The correlations of $\Delta F_4(k,t_1,t_2,t_3)$ at $t_2=0$
between the first time interval $t_1$ and the
subsequent time interval $t_3$ are noticeable over wide time scales.
This result implies that the particle motions are coupled not only at
the (diagonal) $\alpha$- and $\alpha$-relaxation time scales,
but also at the (off-diagonal) $\alpha$- and $\beta$-relaxation time scales.
(iii) With increasing the waiting time $t_2$, $\Delta F_4$ gradually decays
to zero, indicating
that the dynamics change from heterogeneous to
homogeneous because of the memory loss in the correlated motions between
the two
time intervals $t_1$ and $t_3$ for the given waiting time $t_2$.
However, it should be noted that the lineshape of the
NTW model, which exhibits the strong (Arrhenius) glass behavior, differs
from those of the KALJ, WAHN, and SS models, which exhibit
fragile glass (super-Arrhenius) behavior, particularly at $t_2=0$.
We observe the strong correlations of $\Delta F_4$ at small values
of $t_1$ and $t_3$ in the NTW model, which
approximately corresponds to the time scale of the early
$\beta$-relaxation at which $F_s(k, t)$
exhibits damped oscillations on a
plateau as shown in Fig.~\ref{fig_fskt}(d).
This well-known finite-size effect in silica
glasses~\cite{Horbach1996Finite, Coslovich2009Dynamics} would be
smaller if simulations were carried out for larger systems.
Furthermore, 
the relaxation time scale of $\Delta F_4(k,t_1,t_2,t_3)$ depends on the model.
As demonstrated in Fig.~\ref{fig_F4}(d), $\Delta F_4$ of the NTW model decays rapidly.
This time scale is clearly smaller than the $\alpha$-relaxation time determined by the
two-point correlation function.
In contrast, the relaxation of $\Delta F_4$ occurs on a time scale
comparable to (or exceeding) $\tau_\alpha$ in the fragile KALJ, WAHN, and SS models, 
as depicted in Fig.~\ref{fig_F4}(a)-(c).

\begin{figure}[t]
\includegraphics[width=.48\textwidth]{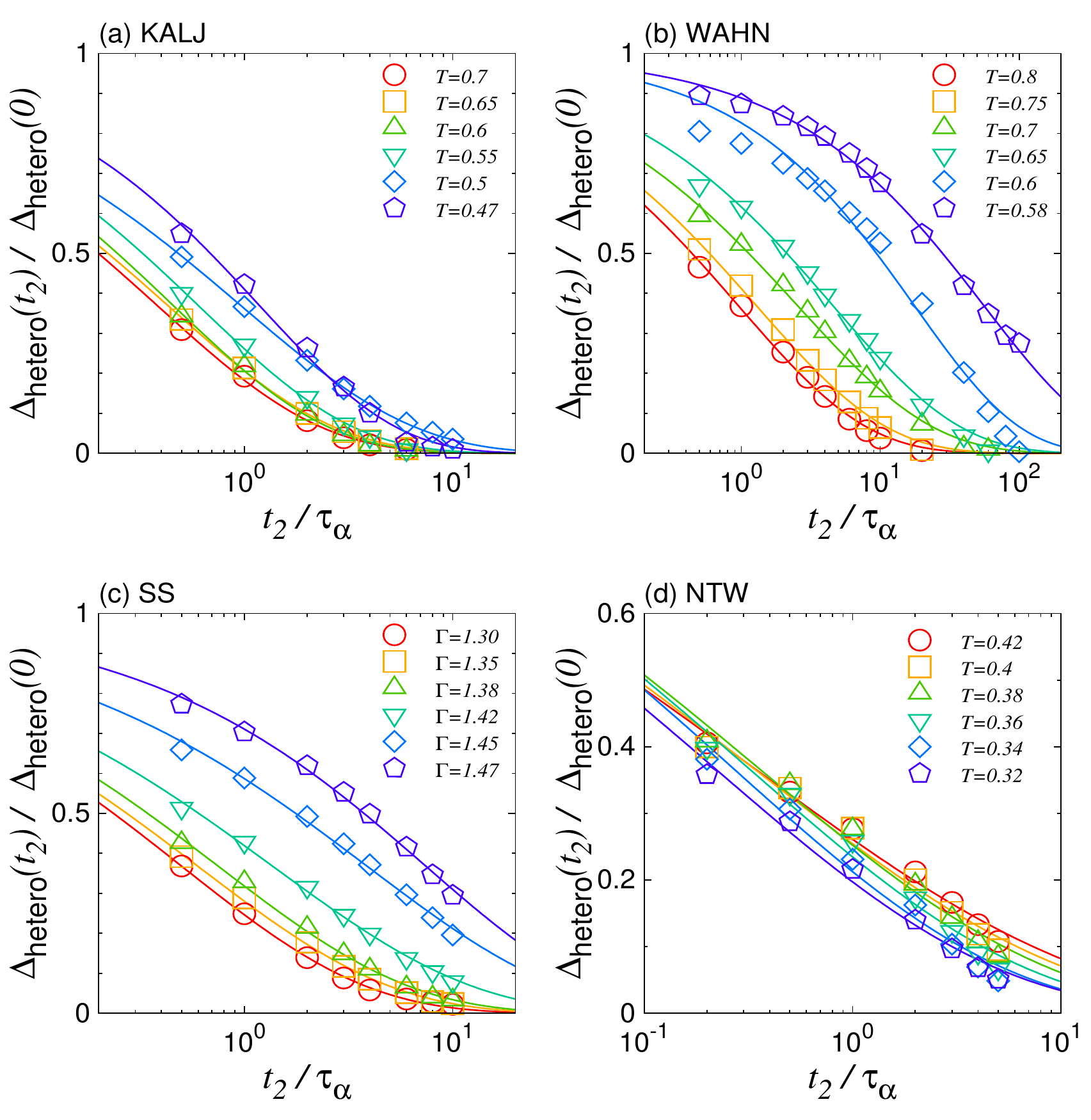}
\caption{
Waiting time $t_2$ dependence of the integrated three-time
 correlation function $\Delta_{\rm hetero}(k, t_2)/\Delta_{\rm hetero}(k, 0)$
for the (a) KALJ, (b) WAHN, (c) SS, and (d) NTW models.
The waiting times are normalized by 
 $\tau_\alpha$ for each temperature.
The solid curve is determined by a fitting with the stretched-exponential form
$\exp[-(t_2/\tau_{\rm hetero})^c]$ with $c\approx 0.6$,
$0.5$, $0.5$, and $0.3$ for the KALJ, WAHN, SS, and NTW models, respectively.
}
\label{sum_hetero}
\end{figure}

To quantitatively distinguish between the models
we quantify the average lifetime of the dynamic heterogeneities 
using the waiting-time $t_2$ dependence of $\Delta F_4$.
To this end, we define the volume of the heterogeneities 
by integrating over the two time intervals $t_1$ and $t_3$~\cite{Kim2009Multiple, Kim2010Multitime},
\begin{equation}
\Delta_{\rm hetero}(t_2) = \int_0^{\infty}dt_3\int_0^{\infty}dt_1 \Delta F_4(k, t_1, t_2, t_3).
\end{equation}
This integration resembles the underlying strategy
of non-linear responses such as NMR, hole-burning, and photo-bleaching
techniques~\cite{Ediger2000Spatially, Wang1999How, Wang2000Lifetime, Schmidt1991Nature, Bohmer1996Dynamic}.
In simulations, similar procedures have been utilized to analyze relevant
multi-time correlations~\cite{Jung2005Dynamical, Leonard2005Lifetime}.
Figure~\ref{sum_hetero} illustrates the $t_2$ dependence of
$\Delta_{\rm hetero}(t_2)$ normalized by $\Delta_{\rm hetero}(0)$ 
for the KALJ, WAHN, SS, and NTW models.
The waiting time $t_2$ is normalized by $\tau_\alpha$ at each temperature.
Figure~\ref{sum_hetero}(d), 
clearly shows that $\Delta_{\rm hetero}$ of the NTW model decays rapidly and
that the relaxation time is smaller than $\tau_\alpha$ at any temperature.
In contrast, Fig.~\ref{sum_hetero}(a)-(c) demonstrates that in the
fragile liquid models, the relaxation of $\Delta_{\rm hetero}$ is slower
than $\tau_\alpha$ with decreasing temperature.
Remarkably, the characteristic time scale of $\Delta_{\rm hetero}$ for
the WAHN model at the
lowest temperature exceeds $\tau_\alpha$ by more than an order of magnitude.

\begin{figure}[t]
\includegraphics[width=.48\textwidth]{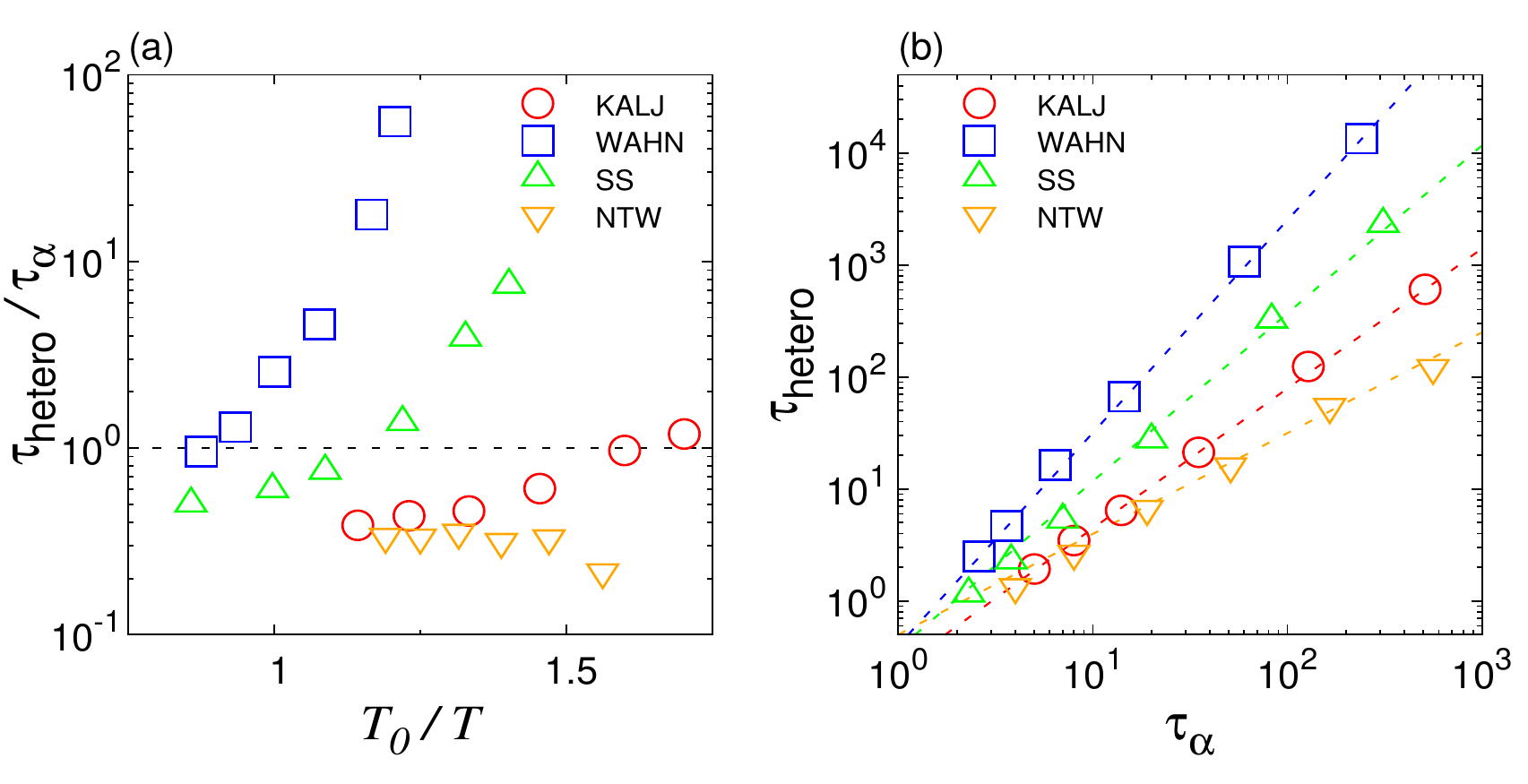}
\caption{
(a) Average lifetime of DH $\tau_{\rm hetero}$ normalized by
 the $\alpha$-relaxation $\tau_\alpha$ as a function of 
 the inverse temperature $T_0/T$ with the onset temperature $T_0$.
The dashed line represents $\tau_{\rm hetero} = \tau_\alpha$ as a viewing guide.
(b) The relationship between two time scales, $\tau_{\rm hetero}$ and $\tau_\alpha$.
The dashed line is the power-law behavior $\tau_{\rm hetero} \sim
 {\tau_\alpha}^{\zeta}$ with a slope of $\zeta\approx 1.9$, $1.5$,
 $1.25$, and $0.9$ from top to bottom.
}
\label{tau_hetero}
\end{figure}

The dependence of the normalized volume
$\Delta_{\rm hetero}(t_2)/\Delta_{\rm hetero}(0)$ on the waiting time
$t_2$ can be fitted by the
stretched-exponential function $\exp[-(t_2/\tau_{\rm hetero})^c]$, as
demonstrated in Fig.~\ref{sum_hetero}.
The exponent $c$ is approximately $c\approx 0.6$,
$0.5$, $0.5$, and $0.3$ for the KALJ, WAHN, SS, and NTW models, respectively.
From this analysis, we determine the average lifetime of the dynamic
heterogeneity as $\tau_{\rm hetero}$ at various temperatures for each model.
We plot $\tau_{\rm hetero}$ as a function of the inverse temperature $T_0/T$ in
Fig.~\ref{tau_hetero}(a).
As shown in Fig.~\ref{tau_hetero}(b), we obtain a
relationship between the two time scales $\tau_{\rm
hetero}$ and $\tau_\alpha$ that follows
the power-law-like behavior,
$\tau_{\rm hetero} \sim {\tau_\alpha}^{\zeta}$ with $\zeta\approx 1.25$,
$1.9$, $1.5$, and $0.9$ for the KALJ, WAHN, SS, and NTW models, respectively.
From the analysis, we find that $\tau_{\rm hetero}$ of the network-forming strong
glass (NTW) is not greater
than $\tau_\alpha$ and tends to decrease with decreasing temperature $T$.
That relationship is responsible for the minor role of the dynamic heterogeneities in
strong liquids, as previously discussed when we examined the four-point
correlations and the associated length scale $\xi$ in Sec.~\ref{four_point}.
In contrast, the ratio $\tau_{\rm
hetero}/\tau_\alpha$ in the fragile liquid models exhibits the opposite
temperature dependence, \textit{i.e.},
the lifetime $\tau_{\rm hetero}$ exceeds the
$\alpha$-relaxation time $\tau_\alpha$ with decreasing temperature.
However, the increase in $\tau_{\rm hetero}$
upon supercooling
is considerably different among the fragile KALJ, WAHN, and SS models.
The ratio $\tau_{\rm hetero}/\tau_\alpha$ markedly exceeds the unity at
lower temperatures in the WAHN and SS models, whereas $\tau_{\rm hetero}$
in the KALJ model remains on the time scale of $\tau_\alpha$, even at
the lowest temperature.
We note that this feature of the KALJ model was also found in a
previous study using a similar multi-time correlation function, in which
its lifetime is comparable to $\tau_\alpha$~\cite{Flenner2004Lifetime}.

A discussion of the significant observed dependence of 
the lifetime $\tau_{\rm hetero}$ on the model details is meaningful,
even for the fragile models.
Recently, the fragility indexes of the simulated models such as the KALJ
and WAHN models 
have been critically investigated from the perspective of the many-body static
correlations hidden in the two-point correlations such as the usual
radial distribution function.
It has been found that the slow- and long-lived correlated domains
correspond to the locally preferred structures (LPSs), that are
characterized by the Voronoi polyhedra~\cite{Coslovich2007Understanding, Coslovich2011Locally}.
These studies have also revealed that the non-additive KALJ mixture is
\textit{less} fragile than
the additive WAHN mixture.
This difference in fragility can be explained in terms of the
spatial extent of the LPS domains.
In fact, the growth of the LPS domains is 
significant in the WAHN model that develops icosahedral order upon supercooling.
In contrast, the LPS domains in the KALJ model formed by a bicapped prismatic
order are found to be smaller than that in the WAHN model.
Given these findings, one can conclude that the overall temperature tendency
of the lifetime $\tau_{\rm hetero}$ shown in Fig.~\ref{tau_hetero}
is correlated and intensely sensitive to the fragility of the model, \textit{i.e.},
more fragile liquids tend to exhibit longer
dynamic heterogeneity lifetimes $\tau_{\rm hetero}$.

\section{Summary}
\label{summary}

We have examined the four-point
correlation function and its three-time correlation extension
to systematically characterize the length and time scales of dynamic
heterogeneities in prototype fragile (KALJ, WAHN, and SS) and strong (NTW)
glass models.
Analyses such as the
extensive investigation performed herein to determine not only the
length scale
but also the time scale of 
various glass-forming models
have not, to the best of our knowledge, been previously reported.

First, we quantified the growing length scale of the dynamic
heterogeneities upon supercooling as determined by the wave-number
dependence of the four-point correlation function using the overlap
function.
The scaling relationships of the extracted length scale $\xi$, which are
analogous to the dynamical scaling obtained for critical phenomena, were
consistently explored 
for the employed glass models.
We observed that the length scale increases with decreasing temperature
depending on the fragility of glass.
In particular, the increase in the dynamic length scale of the strong
glass is suppressed compared with those of the fragile glass-forming liquids,
indicating that the dynamic heterogeneity is less pronounced
and plays only a minor role in the strong liquid.
We also commented on the comparisons of our numerical results
with the theoretical predictions of the IMCT.

Second, we investigated the time scale of the dynamic heterogeneities
from determining
how long the heterogeneous dynamics survive.
Comprehensive numerical results of the three-time
correlation function were demonstrated via two-dimensional correlation maps
with an analogy to the multi-dimensional spectroscopic methods,
as outlined in the introduction.
From the progressive changes in the second time interval of the
three-time correlation function,
we quantified the characteristic time scale of the dynamic
heterogeneities and the associated lifetime
$\tau_{\rm hetero}$.
The lifetime $\tau_{\rm hetero}$ exceeds
the $\alpha$-relaxation time $\tau_\alpha$, particularly for highly supercooled
states in fragile glass-forming liquids.
In contrast, $\tau_{\rm hetero}$ is smaller than $\tau_\alpha$ even at
lower temperatures in the strong liquid, indicating that the dynamic
heterogeneities play a minor role.
Furthermore, we observed that the temperature dependence of
$\tau_{\rm hetero}$ depends significantly on the fragility, \textit{i.e.}, 
more fragile liquids exhibit
long-lived dynamic heterogeneities with a time scale that exceeds the
$\alpha$-relaxation time.
The two time scales differ by more than an order of magnitude in the
WAHN model.

Finally, we remark that it is of important to investigate 
the relationship between the length and time scales of
dynamic heterogeneities and their model dependence.
In fact, we find that 
the length scale $\xi$ of fragile liquids increases with decreasing
temperature in a similar manner, as seen in Fig.~\ref{fig_xi4}(a).
In contrast, 
the time scale $\tau_{\rm hetero}$ is more sensitive to the fragility
and becomes noticeably longer than $\tau_\alpha$ as the fragility index
increases.
To clarify it,
further work that utilizes not only
multi-point and multi-time correlations but
also other measurements including configurational entropy, LPS, PTS, and
BOO is required.

\begin{acknowledgments}
The authors thank Ryoichi Yamamoto, Hideyuki Mizuno, Kunimasa Miyazaki, 
 Takeshi Kawasaki, Hayato Shiba, and Atsushi Ikeda for helpful comments.
We also gratefully acknowledge the information on the NTW model from Daniele Coslovich.
K.K. is grateful to David Reichman and Glen Hocky for valuable discussions.
This work was partially supported by Grant-in-Aid for Young Scientists (A)
 (Grant Number 23684037) (K.K.) and Scientific Research (B) (Grant Number 22350013) (S.S.)
 from Japan Society for the Promotion of Science (JSPS).
This work was also supported by 
the Strategic Programs for Innovative Research (SPIRE) and 
the Computational Materials Science
 Initiative (CMSI)
 of the Ministry of Education, Culture, Sports, Science and Technology, Japan
 (MEXT).
The computations were performed at Research Center of Computational
Science, Okazaki Research Facilities, National Institutes of Natural
 Sciences, Japan.
\end{acknowledgments}

\end{document}